\documentclass[prd,aps,floats,twocolumn,nofootinbib]{revtex4-1}
\usepackage{slashed}
\usepackage{mathtools}
\usepackage{amsfonts}
\usepackage{amssymb}
\usepackage{epsfig}
\usepackage{empheq}

\usepackage{bm}

\begin{document}
\newcommand*\widefbox[1]{\fbox{\hspace{2em}#1\hspace{2em}}}
\newcommand{\m}[1]{\mathcal{#1}}
\newcommand{\nn}{\nonumber}
\newcommand{\ph}{\phantom}
\newcommand{\eps}{\epsilon}
\newcommand{\be}{\begin{equation}}
\newcommand{\ee}{\end{equation}}
\newcommand{\bea}{\begin{eqnarray}}
\newcommand{\eea}{\end{eqnarray}}
\newcommand{\cH}{{\cal H}}
\newtheorem{conj}{Conjecture}

\newcommand{\plk}{\mathfrak{h}}


\title{
Attracting without being attracted: Dark Matter as an aether wind}

\date{}

\author{Raymond Isichei}
\email{raymond.isichei21@imperial.ac.uk }
\author{Jo\~{a}o Magueijo}
\email{magueijo@ic.ac.uk}

\affiliation{Abdus Salam Centre for Theoretical Physics, Imperial College London, Prince Consort Rd., London, SW7 2BZ, United Kingdom}

\begin{abstract}
We explore the possibility that part of what we call dark matter may be the mark of a preferred frame, revealing a breakdown of diffeomorphism invariance. In the non-relativistic limit this appears as a deviant matter source capable of attracting normal matter, but not feeling the attraction from other forms of matter or from itself. While this implies a violation of momentum conservation, no logical inconsistencies arise in this deviant ``Newtonian'' limit. In contrast, due to Bianchi identities, the relativistic theory must undergo core change, and we discuss a modification of Einstein's gravity capable of coupling a non-conserved source to gravity. It results from fixing some of the spatial components of the metric, thereby constraining the possible diffeomorphisms and clipping some of the equations. Bianchi identities can always be used to refill the equations, but the effective Stueckelberg stresses are so outlandish that this defines  symmetry breakdown and violations of local energy-momentum conservation. We work out spherically symmetric solutions with static halos and flat rotation curves, with and without a central black hole. The model has the drawback that it can evade experimental constraints simply by setting to zero the local density of deviant matter (which is a non-dynamic input). Its presence, in contrast, would leave inimitable signatures. We briefly discuss the Hamiltonian formulation of these models, where such dark matter appears as a central charge in the Poisson bracket of the Hamiltonian and the momentum.


\end{abstract}

\maketitle

\section{Introduction}

For all its successes, the dark matter paradigm cannot fend off the criticism that what we call dark matter might be a mere phenomenological proxy for a more fundamental theoretical construction.
The model is also not without observational shortcomings and epicycles. A case in point is galactic-scale  dark matter. Dark matter 
explains observed gravitational pull by means of  non-observed ``dark'' gravitational sources, 
but, for consistency, the dark sources should stay where they are needed under their own gravitational pull and that of visible forms of matter. 
This does not always happen without contortions, for example in the case of galactic halos (e.g. fuzzy dark matter~\cite{fuzzy}, triaxial bars~\cite{triaxial}, among others).

None of this caused undue consternation at the inception of the dark matter paradigm; indeed  
the {\it ad hoc} phenomenological nature of dark matter was initially cherished without shame. At its most extreme (and possibly facetiously), the idea of  ``painted-on'' dark matter was informally floated: a {\it subdominant} dust-like component capable of attracting visible matter, but staying put where it was required (hence the moniker``painted on''), oblivious to its own gravity or that of other sources. Regardless of the phenomenological merits of this extremist view, it will provide the broad inspiration for this paper.  
Could such subdominant painted-on dark matter be encrusted in, and so be the tell-tale signature of a preferred frame in the Universe? There is nothing more fundamental than deciding whether such a frame exists: this is a throwback to the old aether quest. Combining such a foundational issue with dark matter could therefore be a step forward in understanding its true nature.


Implementing such an idea in the context of General Relativity, however, is far from trivial. 
Within the Newtonian perspective/limit there is {\it a priori} nothing inconsistent with an object attracting but not being attracted. It contradicts Newton's third law and so leads to violations of momentum conservation, but no logical inconsistency arises. Indeed, at the Newtonian level, painted-on dark matter can be introduced by directly positing a dust-like fluid with active gravitational mass ($m_a\neq 0$) but no passive gravitational mass ($m_p=0$), or alternatively to the latter, infinite inertial mass ($m_i=\infty$). Such a fluid would produce gravity according to the usual Newton law of gravity, but feel no force of gravity if $m_p=0$, or fail to be budged by any force if $m_i=\infty$.

The situation is dramatically different in General Relativity, because concepts such as inertial mass and (active and passive) gravitational mass vanish, and are
traded for geodesic motion for point particles.
Crucially, such geodesic motion is not a postulate, but is implied by stress-energy tensor conservation, itself automatically enforced by the Bianchi identities for the Einstein tensor $G_{\mu\nu}$ and the gravitational field equations. Inserting painted-on dark matter (or any other matter source violating local energy-momentum conservation) into the Einstein equations leads to a logical contradiction akin to $1=2$. 

There are several attitudes one can take towards this. One is to accept that an entirely different theory of gravity needs to be built to accommodate a non-conserved source. Another is to
ask whether General Relativity can be minimally modified to accommodate such an alien concept. In this paper we take the latter route, skirting the perils of trivial solutions. If we do not radically depart from General Relativity, the theory wants to force on us a solution effectively amounting to a standard conserved source, e.g. an equivalent of fuzzy dark matter. This threat will always be breathing down our necks throughout this paper, but will be avoided by making such equivalent completions, reinstating conservation, sufficiently ludicrous. This is closely related to the Stueckelberg method for reinstating symmetry and will be discussed later in this paper.  

A bit of background. Our approach follows from a series of papers where the Hamiltonian constraint is violated in some regimes (high energy, early Universe, etc) only to be restored in others. Invariably such symmetry restoration cannot avoid an indelible memory of the broken phase in the form of something that looks like, but is not dark matter. This happens in Horava-Lifshitz theory~\cite{HL,shinji}, in theories where there is evolution (or time dependence) in the laws of physics~\cite{geoCDM,nongeoCDM}, or as an outcome of quantization~\cite{StueckelDM,Barvinsky,kaplan1,kaplan2}. A passing connection with mimetic dark matter exists~\cite{mimetic}. Finally, the fact that this effect can result from the interaction between a global, presiding Universe and local physics led to this component  being called ``Machian'' dark matter in~\cite{MachianCDM}. 

We acknowledge all of this past work, but stress that 
the novelty here is that we are proposing a permanent breakdown of diffeomorphism invariance wherever such violations of the Hamiltonian constraint took place.
An analogy (in reverse) with topological defects will be drawn in the concluding remarks of this paper. 

Throughout this paper we will use units such that $G=1$ (rather than $8\pi G=1$).


\section{Non-conserved sources and what they impose on gravity}
Given our intention to abandon the redundancy of local energy-momentum conservation for ``normal matter'' due to Bianchi identities, we  start by imposing it as a separate postulate:
\begin{equation}\label{Mcons}
    \nabla_\mu T_M^{\mu\nu}=0.
\end{equation}
We then add to normal matter a component of non-conserved matter (additively, so that the total is $T_{\mu\nu}=T_\Delta^{\mu\nu}+T_M^{\mu\nu}$). In this paper we choose a component with energy-momentum tensor of the form:
\begin{equation}\label{TmnDelta}
    T_\Delta^{\mu\nu}=\frac{\Delta(x)}{\sqrt{h}}n^\mu n^\nu,
\end{equation}
where $x$ is a 3D spatial coordinate on the preferred foliation $\Sigma_t$, and $n^\mu$ is the normal (with ADM conventions~\cite{MTW,ADMReview}) to the preferred frame, not necessarily geodesic. 
Importantly, $\Delta(x)$ is non-dynamical: it is ``painted-on'' the preferred frame. The reason for this choice is that it connects with historical/high energy violations of the Hamiltonian constraint~\cite{shinji,geoCDM,nongeoCDM,StueckelDM,kaplan1,mimetic,MachianCDM}, with $\Delta(x)$ representing the value of the leftover Hamiltonian, once the driving mechanism switches off. The connection with the Hamiltonian framework will be explored in greater detail in Section~\ref{HamSec}.

We see that:
\begin{align}
    n_\mu\nabla_\nu T^{\mu\nu}_\Delta & =\frac{1}{\sqrt{-g}}\partial_i(N^i \Delta)\label{violE}\\
     h_{i\mu}\nabla_\nu T^{\mu\nu}_\Delta & =\frac{1}{\sqrt{-g}}\Delta \partial_iN.\label{violP}
    \end{align}
This may be proved directly, or else using the Hamiltonian framework, as was done in~\cite{geoCDM,nongeoCDM} (see also \eqref{equiv1} and \eqref{equiv2} later in this paper).
The question is what can the gravitational field equations be, given these matter sources?
We may consider a left hand side with $H_{\mu\nu}\neq G_{\mu\nu}$ but closely related to $G_{\mu\nu}$ (for example preserving some of its components but not others), so that the full set of independent equations reads:
\begin{align}\label{Hmunu}
    H^{\mu\nu} &= 
    8\pi (T_\Delta^{\mu\nu}+ T_M^{\mu\nu}).\\
     \nabla_\mu T_M^{\mu\nu}&=0.
\end{align}
This requires that $H_{\mu\nu}$ violate Bianchi identities accordingly:
\begin{align}
    n_\mu\nabla_\nu H^{\mu\nu} & =\frac{8\pi }{\sqrt{-g}}\partial_i(N^i \Delta)\label{noncon1}\\
     h_{i\mu}\nabla_\nu H^{\mu\nu} & =\frac{8\pi }{\sqrt{-g}}\Delta \partial_iN.\label{noncon2}
    \end{align}
There is an infinite number of such $H_{\mu\nu}$ and, as already flagged in the Introduction, we must avoid the trivial ones: those that simply result in $G_{\mu\nu}$ plus terms that effectively endow $T_\Delta^{\mu\nu}$ with standard local dynamical variables rendering it a conserved matter source. In such a case we would be back to square one: fuzzy dark matter and the like, in the context of galactic halos. As respectable as these approaches are, that is not the purpose of this paper. 



\section{Clipping}\label{Sec:clippersGen}
A possible solution is to decrease the number of Einstein equations by ``clipping'' some of them, reducing in tandem the number of metric variables, to keep the system well defined. This can break the contradiction between a non-conserved source and the Bianchi identities. In general the new field equations will be:
\begin{align}
     P_{\mu\nu}^{\;\;\;\alpha\beta}[G_{\alpha\beta}-8\pi (T^\Delta_{\alpha\beta}+T^M_{\alpha\beta})]&=0\label{clipped}\\
      \nabla_\mu T_M^{\mu\nu}&=0\label{Mcons-pair}
\end{align}
where $P_{\mu\nu}^{\;\;\;\alpha\beta}$ projects out, or ``clips'' some components. In general the projector takes the form:
\begin{equation}
     P_{\mu\nu}^{\;\;\;\alpha\beta}=\delta_{\mu}^\alpha\delta _\nu^{\beta}\pm N_{\mu\nu}N^{\alpha\beta}
\end{equation}
(the sign depending on the nature of the clipping),
with $N^{\mu\nu}$ encoding the components to be clipped. 


The process of clipping could arise from constraining the metric {\it before} variations of the Einstein-Hilbert action are taken.  Let the conditions (indexed by $i$) upon the metric  take the general form:
\begin{equation}\label{freezemetric}
    F_i(g_{\mu\nu})=0
\end{equation}
which need not be covariant and could involve differential operators.  These conditions can be seen as the result of removing some diffeomorphisms, specifically for $x^\mu\rightarrow x^\mu +\xi^\mu$, with:
\begin{equation}\label{freezediffeo}
    F_i(\xi_{\mu;\nu})=0.
\end{equation}
This is similar but fundamentally different from gauge fixing, in that it states that the corresponding gauge symmetry never existed due to $\Delta$. Furthermore, we insist on preserving full symmetry for normal matter, so that \eqref{Mcons} remains valid.

Such clipping is the most conservative option (as in the one closest to the spirit of General Relativity). 
But we could also clip an Einstein equation arising from varying the action with respect to a different metric component than the one we have frozen. This is off-piste and therefore interesting. It would require a revision or even a definite abandonment of the action principle formulation. We will present one such example in Appendix~\ref{Sec:radial}.


Such ``clippings'' are not new. For example, unimodular gravity 
~\cite{unimod1,unimod,UnimodLee1,unimod,alan,daughton,sorkin1,sorkin2,Bombelli,UnimodLee2} can be built from:
\begin{equation}
    P_{\mu\nu}^{\;\;\;\alpha\beta}=\delta_{\mu}^\alpha\delta _\nu^{\beta}-\frac{1}{4}g_{\mu\nu}g^{\alpha\beta}
\end{equation}
(one does not vary with respect to the trace of the metric),
whereas projectable Horava-Lifschitz theory~\cite{HL,shinji} and similar theories of evolution~\cite{geoCDM} employ:
\begin{equation}\label{clippergeoCDM}
    P_{\mu\nu}^{\;\;\;\alpha\beta}=\delta_{\mu}^\alpha\delta _\nu^{\beta}-n_\mu n_\nu n^\alpha n^\beta
\end{equation}
(one does not vary with respect to a local lapse function defined on the preferred frame $\Sigma_t$). For a non-projectable version of the theory~\cite{nongeoCDM}, the momentum constraint is also removed:
\begin{equation}
    P_{\mu\nu}^{\;\;\;\alpha\beta}=\delta_{\mu}^\alpha\delta _\nu^{\beta}-n_{(\mu} h_{\nu )}^\gamma  n^\alpha h_\gamma^\beta.
\end{equation}
In these known examples we find that, having {\it clipped} the Einstein equations, we can {\it refill} them using the Bianchi identities and \eqref{Mcons}, with the procedure adding to the equations an ``integration constant'' $S_{\mu\nu}$:
\begin{equation}\label{Smunu}
    G_{\mu\nu}=8\pi (T_{\mu\nu}+ S_{\mu\nu})
\end{equation}
(where $T_{\mu\nu}$ is the total stress-energy tensor, containing normal matter and $\Delta$). The new $S_{\mu\nu}$
satisfies:
\begin{align}
    P^{\;\;\;\alpha\beta}_{\mu\nu}S_{\alpha\beta}&=0\\
      \nabla_\mu S^{\mu\nu}&= -  \nabla_\mu T^{\mu\nu}\label{divST}
\end{align}
This ``integration constant'' is the cosmological constant $\Lambda$ in the first case, a dust like fluid in the second, and the unusual fluid described in~\cite{nongeoCDM} in the third case. In all these cases \eqref{divST} vanishes: as already stressed in the Introduction, this is one aspect in which painted-on matter is different from previous literature. We can interpret $S_{\mu\nu}$ as the Stueckelberg fields restoring full diffeomorphims invariance (for this interpretation regarding unimodular gravity see~\cite{Saltas}), and this will be critically appraised in our case later in this paper.



\section{Clippers for painted-on matter}\label{Sec:clippers}
Our case is different in that we start from a pre-given non-dynamical $\Delta(x)$ on $\Sigma_t$ (and so, $\Delta$ cannot be an integration constant due to refilling), implying that we cannot clip the Hamiltonian constraint (as in~\cite{shinji,geoCDM}), which is the only place where $\Delta$ appears. Hence:
\begin{equation}
     P^{\;\;\;\alpha\beta}_{\mu\nu}T^\Delta_{\alpha\beta}=T^\Delta_{\alpha\beta}
\end{equation}
ruling out a term of the form \eqref{clippergeoCDM}. Instead, clipping is introduced in our case to remove the algebraic inconsistency arising from the non-conservation of $T^\Delta_{\alpha\beta}$ and the Bianchi identities for the unclipped Einstein equations. 

Furthermore, since $\Delta$ is a non-dynamical ``given'', it does not need to be the most general field, that is, we can impose any conditions on $\Delta$ that we wish, each case corresponding to a different theory. For example we could impose that it must be a constant or piece-wise constant\footnote{Note that this condition explicitly breaks spatial diffeomorphism invariance, since $\Delta$ is a density, and a constant density is not covariantly constant. The statement is pegged to preferred 3D coordinate frames.},
with regions of constant $\Delta\neq 0$ (in some spatial coordinates) acting as a ``defect'' in diffeomorphism invariance. This would give standard CDM if implemented on a cosmological scale (\cite{shinji,geoCDM}; the cosmological frame is geodesic). On a particular non-geodesic frame it would be something akin to painted-on galactic halos, as explored in Section~\ref{SStesttube}, particularly~\ref{halocritical}. This is a possibility and an interesting example, but many other options could be explored. At the most extreme, we could even require that the regions with constant $\Delta\neq 0$ be spherical. 

Bearing these two important points in mind we then note that not all clippers enforce consistency. For example, building the clipper from tensor products of $D_i\Delta$ would not work, since a covariantly constant $\Delta$ can lead to violations. Reciprocally, there is typically more than one set of clippers that can enforce consistency (for a given $\Delta$). Each choice leads to a different theory. 



For the sake of concreteness, we will explore one possible clipping mechanism (and defer to Section~\ref{HamSec} and further work, the Hamiltonian recipes for clipping). Face value, there are two degrees of freedom in the violations of local energy conservation~\eqref{violE} and~\eqref{violP}, so one might think that two Einstein equations need to be clipped\footnote{This may be overkill, as the Hamiltonian perspective shows, but that is not the point here.}.
As we will see, \eqref{violE} does not appear in the Newtonian limit or in the leading Post-Newtonian terms. In addition it can be dealt with by fixing one of the shift functions $N^i$ such that $\partial_i(N^i\Delta)=0$. This restricts diffeomorphisms by:
\begin{equation}\label{diffreduct}
    \partial_i[\Delta (x)N^2 (\xi ^{0;i} +N^i\xi^{0;0})]=0
\end{equation}
and this procedure, by refilling leads to the fluid in \cite{nongeoCDM}. This is not where the novelty in this paper lies.

Let us focus instead on \eqref{violP}. Given that we are breaking 4D diffeomorphism invariance, we can clip with purely spatial structures on $\Sigma_t$ (that is, clip spatial Einstein equations, or equations that result from Hamilton's equations, rather than the constraints, as we have done for \eqref{violE}). An obvious structure on $\Sigma_t$ is formed by the iso-lapse surfaces, that is, the 2D surfaces $\Sigma_N$ of constant lapse function $N$. (Note that if $N$ is constant on $\Sigma_t$, there are no violations \eqref{violP}, the frame is geodesic and this is just~\cite{shinji,geoCDM}.) The normal to such 2D spatial surfaces $\Sigma_N$ is determined by the spatial part of the acceleration of $\Sigma_t$, $a_\mu =n^\alpha\nabla_\alpha n_\mu$, with $a_i=\partial_i \log N$. 
The clipper should thefore employ a normalized version of this acceleration:
\begin{equation}
    u_\mu\equiv \frac{a_\mu}{\sqrt{a_\mu a^\mu}}.
\end{equation}
We can also do a 2+1 ADM decomposition based on surfaces
$\Sigma_N$ with lapse $M$, defining the normal $u_\mu=(-M,0)$. 

We can now define a variety of clippers based on $u_\mu$. 
Each prescription leads to a different theory. In analogy with the unimodular cosmological constant, we could for example employ a transverse clipper:
\begin{equation}
     P_{\mu\nu}^{\;\;\;\alpha\beta}=\delta_{\mu}^\alpha\delta _\nu^{\beta}-\frac{1}{2}\prescript{(2)}{}h_{\mu\nu}\prescript{(2)}{}h^{\alpha\beta}
\end{equation}
with $\prescript{(2)}{}h_{\mu\nu}=g_{\mu\nu}+n_\mu n_\nu-u_\mu u_\nu$ the induced metric on the surfaces of constant $N$. Or we could use a longitudinal clipper
\begin{equation}
     P_{\mu\nu}^{\;\;\;\alpha\beta}=\delta_{\mu}^\alpha\delta _\nu^{\beta}-u_\mu u_\nu   u^\alpha u^\beta
\end{equation}
or a combination of the two, or even more complex structures if need be. In this paper 
we are not wedded to any of these theories but would rather explore the lot of them. 

Returning to the second point made at the start of this Section, we stress that there is no a priori reason why such clippers would enforce consistency for the most general $\Delta$, but as pointed out, $\Delta$ does not need to be general. We can always impose a condition upon $\Delta$ such that each of these clippers is consistent.


\section{The futility of ``refilling'' for painted-on matter}\label{futility}

The case in hand is also different (compared to other clippings found in the literature) in that ``refilling'' now produces not $T^\Delta_{\mu\nu}$, but rather the relevant momentum and stresses to render it conserved. The ``integration constant'' $S_{\mu\nu}$ in \eqref{Smunu} 
(the analogue of $\Lambda$ in unimodular gravity)
is not $\Delta$ itself, but the spatial pressures and anisotropic stresses (and, to second order, the momentum as in~\cite{nongeoCDM}) required to render it conserved. 
This is an expression of the {\it superficial} power of Bianchi identities. If we start by accepting some components of the Einstein tensor and clip others (fewer than 4), nothing can stop us, modulo technicalities,  from refilling the latter by integrating the Bianchi identities, resulting in an ``integration constant'' to be inserted in the refilled Einstein equations.
The ``integration constant'' can be seen as the Stueckelberg field restoring full diffeomorphism invariance, and in our case stress-energy-momentum conservation.

But this can be misleading and vacuous. Indeed, by making diffeomorphism invariance (and local energy-momentum conservation) {\it always} possible, the Bianchi-Stueckelberg procedure has become content-free by itself. The situation is similar to the Einstein/Friedrichs debate~\cite{Fried}. Einstein used general covariance as the guiding principle leading to General Relativity; however any theory can be made generally covariant, including Galilean theory, as shown in~\cite{Fried}. The difference is that some theories (such as Galilean theories) require the introduction of forceful and artificial structures (effectively the Stueckelberg fields adapted to Galilean theories). The absence of such structures is what sets apart General Relativity and other truly diffeormorphism invariant'' theories.

Likewise, in our context, we say that a theory has violations of diffeomorphism invariance and of local energy-momentum conservation, if the $S_{\mu\nu}$ that return the theory to conservation are forceful/artificial, either by satisfying a non-local dynamics, or by breaking any reasonable energy condition, or by implying stresses depending on all other fluids and global environment (rather than a traditional equation of state, where the stresses would be local functions of $\Delta(x)$), or any such ``preposterous'' property. 
In our case, for a projector $ P^{\;\;\;\alpha\beta}_{\mu\nu}$, the Stueckelberg pressures $S^{\mu\nu}$ must satisfy:
\begin{align}
    P^{\;\;\;\alpha\beta}_{\mu\nu}S_{\alpha\beta}&=0\label{Stueckel1}\\
    \nabla_\mu S^{\mu\nu}&=-\nabla_\mu T_\Delta^{\mu\nu}
    \label{Stueckel2}
\end{align} 
with decompositions \eqref{noncon1} and \eqref{noncon2} for the right hand side of \eqref{Stueckel2}. Hence, 
$S_{\mu\nu}$ depends non-locally on all types of matter, not just locally on $\Delta$ (or else it can depend locally on $\Delta$ via a severely non-minimal coupling to gravity).  This is not an equation of state as for a fluid, nor a local dynamics as for a scalar field. In fact the general form of the equations for the projectors discussed above is so cumbersome that we will not present them here (even though we will discuss their form in the spherically symmetric cases in Section~\ref{SStesttube} and Appendix~\ref{Sec:radial}). 

Note that using the tensor defined in \eqref{Hmunu}, the field $S_{\mu\nu}$  reads (or is defined as):
\begin{equation}
    H_{\mu\nu}=G_{\mu\nu}-8\pi S_{\mu\nu}.
\end{equation}

\section{Newtonian and Post-Newtonian views}
As already pointed out in the introduction, in the Newtonian theory there is nothing self-contradictory about a source that attracts but is not attracted (with consequent momentum conservation violation). This can be seen in the Newtonian limit of a relativistic theory using the Post-Newtonian (PN) expansion: 
\begin{align}
		N &= 1 + \frac{\phi}{c^2} + {\cal O}(  c^{-4} )\\
		h_{ij} &= \gamma_{ij}\left(1 - \frac{2\psi}{c^2}\right) + {\cal O}( c^{-4} ) 
\end{align}
(where $\gamma_{ij}$ is the unperturbed flat metric in a generic coordinate system) leading to:
\begin{align*}
	G^{00} &= \frac{2}{c^2}\Delta\psi + {\cal O}( c^{-4} )\,,
	\\
	G^{0i} &= \frac{2}{c^3}\left[ \Delta\zeta^i -  D^i D_j \zeta^j - D^i\dot{\psi}\right] + {\cal O}( c^{-5} ),
	\\
	G^{ij} &= -\frac{1}{c^2}\left[ \Delta\left(\phi-\psi\right)\gamma^{ij} - D^i D^j  \left(\phi-\psi\right)\right] + {\cal O}( c^{-4} )\,.
\end{align*}
As we can see by inspection, 
there is no Bianchi identity to lowest order ($1/c^2$), so the Newtonian solution:
\begin{align}
    \nabla\psi&=4\pi \Delta\\
    \phi&=\psi
\end{align}
is fully unconstrained.
The lowest order Bianchi identities are of order $1/c^3$ (for $\dot G^{00}$) and to order $1/c^4$ (for $\dot G^{0i}$). They imply conditions upon the higher order momentum and spatial stresses enforcing energy-momentum conservation. 
Their expression for the clippers proposed above is very complicated and makes the point that they are artifical, but the issue can be understood qualitatively.


Consider for example momentum conservation for a blob
of $\Delta(x)$ initially on its own. 
A normal matter source approaches, attracted to it, but the $\Delta$ blob does not budge, feeling no attraction to it. This can be seen as a violation of momentum conservation. It can also be reinterpreted as evidence of new stresses/pressure acting on the blob to counteract the gravitational forces of normal matter, with energy momentum conservation.  However, the stresses depend on other forms of matter rather than on $\Delta(x)$ and worse of all the dependence is non-local, since they must counter the Newtonian force generated by the whole (normal) matter source. The stresses required to make $T^{\mu\nu}_\Delta$ conserved can never arise from an equation of state. 




\section{Spherically symmetric solutions with transverse clipping}\label{SStesttube}
Let us now consider a static spherically symmetric space-time with both $\Delta$ and normal matter, which takes the form of a fluid with equation of state:
\begin{equation}
      p_M=p_M(\rho_M).\label{eqstat}
\end{equation}
We apply to the theory a transverse clipper (and leave the case of a radial clipper to Appendix~\ref{Sec:radial}). Let the conditions associated with the clipper force the 2D metric transverse to the iso-$N$ lines to be the Euclidean metric in spherical coordinates. Then, the metric is:
\be
ds^2 = -e^{2\Phi(r)} \, dt^2 + e^{2\lambda(r)} \, dr^2 + r^2 d\Omega^2\nn
\ee
where $d\Omega^2= d\theta^2 + \sin^2 \theta \, d\phi^2 $. In addition 
$\Delta=\Delta(r)$, implying:
\begin{align}\label{rhoDelta}
    \rho_\Delta&=\frac{\Delta(r)}{r^2 e^\lambda}
\end{align}
where, we recall, $\Delta$ is a 3D scalar density. 
We stress that we are {\it not} using gauge freedom to choose this metric (as in standard General Relativity textbook treatments), but instead freezing the metric angular variables at the level of the action. Hence, there are no angular Einstein equations, as a particular case of \eqref{freezemetric} (and disallowing diffeomorphisms as in \eqref{freezediffeo}). The equations of motion \eqref{clipped} and \eqref{Mcons-pair} under spherical symmetry with a transverse clipper therefore read:
\begin{align}
    e^{-2\lambda} \left[ \frac{2}{r} \lambda' - \frac{1}{r^2} \right] + \frac{1}{r^2} &= 8 \pi (\rho_\Delta+\rho_M),\label{G00T}\\
    \left[ \frac{2}{r} \Phi' + \frac{1}{r^2} \right] e^{-2\lambda} - \frac{1}{r^2} &= 8 \pi p_M, \label{GrrT} \\
    -\Phi'(\rho_M+p_M) &=p_M',\label{MConsT}
\end{align}
(where, we recall, we have adopted $G=1$). 
Just as with the Tolman-Oppenheimer-Volkov (TOV) system, by setting:
\begin{equation}\label{lambdamrT}
    e^{-2\lambda}=1-\frac{2m}{r}
\end{equation}
equation \eqref{G00T} gives:
\begin{equation}\label{mprimeT}
    m'=
    4\pi \rho r^2= 4\pi\left(
    \rho_M r^2+\Delta(r) \sqrt{1-\frac{2m}{r}}\right),
\end{equation}
with $\rho=\rho_M+\rho_\Delta$.
Eq.~\eqref{GrrT} then implies:
\begin{equation}\label{TOVphiprime}
     \Phi' = \frac{m + 4\pi r^3 p_M}{r(r - 2m)},
\end{equation}
so that if $\Delta=0$ we do recover the original 3 TOV ordinary differential equations, which together with the equation of state \eqref{eqstat} can be solved for the 4 variables $m(r)$, $\rho_M(r)$, $p_M(r)$ and $\Phi(r)$ (with $\lambda(r)$ given separately by \eqref{lambdamrT}).  Since $\Delta$ is {\it not} a variable, by adding it to the system we do not change its nature (i.e. the number of equations and variables). 

However, adding the (usually redundant) angular/transverse equation:
\begin{align}
       e^{-2\lambda} \left[ \Phi'' + \Phi'^2 - \Phi' \lambda' + \frac{\Phi'}{r} - \frac{\lambda'}{r} \right] &= 8 \pi p_M,\label{GangularT}
\end{align}
would render the system self-contradictory. This is the purpose of clipping. 

\subsection{The critical halo}\label{halocritical}
It is curious that a constant\footnote{We again note that such $\Delta$ is ``constant'' as opposed to ``covariantly constant''; recall that it is a tensor density. But since we are breaking covariance, this can hardly be used as an argument for what should be ``naturally'' constant in a uniform distribution of ``matter''. Perhaps tensor densities survive the breakdown of diffeomorphism invariance better than traditional tensors.} $\Delta=\Delta_0$ (with a cutoff $r<r_{\rm max}$) leads to a halo with the correct profile to generate flat rotation curves. We call this profile the ``critical halo''. In the absence of normal matter ($\rho_M=p_M=0$), \eqref{lambdamrT} and \eqref{mprimeT} imply $m=m_0 r$ with: 
\begin{equation}
    \Delta_0=\frac{m_0}{4\pi \sqrt{1-2m_0}}
\end{equation}
and $\lambda=\lambda_0$, with:
\begin{equation}
    e^{-2\lambda_0}=1-2m_0.
\end{equation}
We are assuming that $m_0<1/2$, so that the halo is everywhere outside its Schwarzchild radius. The energy density \eqref{rhoDelta} for this profile is:
\begin{equation}
    \rho_\Delta=
    \frac{\Delta}{r^2 e^\lambda} = 
    \frac{m_0}{4\pi r^2}.
\end{equation}
Then, \eqref{TOVphiprime} leads to the Newtonian potential:
\begin{equation}\label{logPhi}
    \Phi=\frac{m_0}{1-2m_0}\ln\frac{r}{r_c}
\end{equation}
so that the metric is:
\begin{equation}\label{halometric}
    ds^2=-\left(\frac{r}{r_c}\right)^{\frac{2m_0}{1-2m_0}}dt^2+\frac{dr^2}{1-2m_0}+r^2d\Omega^2.
\end{equation}
The solution outside the halo, $r>r_{\rm max}$, is the Schwarzchild solution for mass $M=m_0r_{\rm max}$. Hence,  the integration constant $r_c$ should be adjusted to match $\Phi$ on both sides of $r=r_{\rm max}$. The above solution is exact, but one can take its Newtonian limit and prove that it  leads to flat rotation curves, as clearly implied by \eqref{logPhi}.

In the above, the halo starts from  $r=0$, but it could start at any $0<r<r_{max}$, with similar results. If the halo does start at $r=0$, then there is a naked singularity at $r=0$ (the scalar $\rho_\Delta$ diverges). There are no horizons in this solution. 


\subsection{A critical halo around a black hole}\label{BHhalo}
Adding normal matter does not qualitatively change our results in the Newtonian limit, but beyond that we see that we lose superposition, given that we have to solve the differential equation \eqref{mprimeT} (which is no longer a first integral). Even the critical halo would stop being as simple as in Section~\ref{halocritical} if normal matter were added on.
As a pedantic exercise, 
we could consider a star of normal matter and a critical halo, and study the Oppenheimer-Volkov limit. 

A more relevant problem is the  extreme case where a black hole of normal matter forms, but the cut off at $r<r_0$ for the halo is inserted to prevent the existence of a naked singularity pior to black hole formation.  Throughout this paper we have assumed a preferred space-like surface $\Sigma_t$, so, if a black hole does form
it cannot eat the $\Sigma_t$ and all its leaves pile up at the horizon (a phenomenon already studied in~\cite{BHevol}). 
A possible model follows from:
\begin{align}
    m=M&\qquad{\rm for}\quad  r<r_0\\
    m=M+m_0(r-r_0)&\qquad{\rm for}\quad  r>r_0
\end{align}
with $r_0>2M$ as the simplest choice\footnote{Starting the halo from $r_0=0$ would simply shift $M$ in what follows,
but note that $r$ becomes time-like inside the horizon.}.

For $r<r_0$ the metric is Schwarzchild with mass $M$. For $r_0<r<r_{max}$ integrating \eqref{mprimeT} and \eqref{TOVphiprime} leads to: 
\begin{align}\label{BHhalometric}
    ds^2=&-\frac{[r-2(M+m_0(r-r_0))]^\frac{1}{1-2m_0}}{r(r_0-2M)^\frac{2m_0}{1-2m_0}}dt^2\nn\\&
    +\frac{dr^2}{1-\frac{2}{r}(M+m_0(r-r_0))}+r^2d\Omega^2.
\end{align}
The match point $r_0$ could be anywhere, but it is natural to choose it just outside the horizon $r_0=2M^+$, 
since it is where the $\Sigma_t$ pile up. 

This is an interesting black hole plus critical halo solution.   
One may object that there is an element of teleology in keeping the $\Sigma_t$ outside the black hole's horizon, but there is already teleology in the definition of the concept of horizon.

\subsection{The cost of refilling for these solutions}\label{refillSS}
We now illustrate the points on ``refilling'' made in Section~\ref{futility}, with reference to the solutions found in this Section. Given \eqref{Stueckel1} the only non-vanishing components of the Stueckelberg fields  $S_{\mu\nu}$ are those corresponding to a transverse pressure:
\begin{align}
    p^\Delta_\perp=\frac{S_{\theta\theta}}{r^2}=
     \frac{S_{\phi\phi}}{r^2\sin ^2\theta}.
\end{align}
This can be found either by evaluating the right hand side of  \eqref{Stueckel2} and integrating it, or, equivalently, by replacing  
\begin{equation}\label{peffT}
p_M\rightarrow p_M+ p^\Delta_\perp
\end{equation}
in the clipped equation \eqref{GangularT}, so that:
\begin{align}\label{PeffeqT}
    p^\Delta_\perp=-p_M +\frac{e^{-2\lambda}}{8\pi} \left[ \Phi'' + \Phi'^2 - \Phi' \lambda' + \frac{\Phi'}{r} - \frac{\lambda'}{r} \right] 
\end{align}
with $\Phi$ and $\lambda$ evaluated with the clipped equations, that is, as above. Either way we find:
\begin{equation}
    p^\Delta_\perp=\frac{r}{2}\rho_\Delta\Phi'=\frac{r}{2}\rho_\Delta\frac{4\pi\int dr^2\rho + r^3 p_M}{r(r - 8\pi\int dr^2\rho)}
\end{equation}
(where $\rho=\rho_M+\rho_\Delta$, we recall). As we see (and as announced in Section~\ref{futility}) this is far from an equation of state (such as \eqref{eqstat} for normal matter). The Stueckelberg pressure depends not only on $\Delta$ but also non-locally on all other forms of matter. 
For a pure critical halo we have the accidental relation
\begin{equation}
    p^\Delta_\perp=\frac{m_0}{2(1-2m_0)}\rho_\Delta
\end{equation}
but this is not an equation of state (indeed it varies from halo to halo and the relation evaporates altogether as soon as normal matter is added on).

Regarding the black hole plus halo solution in Section~\ref{BHhalo}, we further note that there are no divergences at $r=2M$ in the clipped theory, since the components of the Einstein tensor that would have diverged have been replaced by $H_{\theta\theta}=H_{\phi\phi}=0$. If we refill, i.e. restore diffeomorphism invariance with Stueckelberg fields $S_{\mu\nu}$, then these do diverge, with:
\begin{equation}
p^\Delta_\perp\rightarrow
    \frac{m_0(1+2m_0)}{2(r-2M)},
\end{equation}
as $r\rightarrow 2M$. Thus, in this reinterpretation there is a pressure singularity at the horizon, not dissimilar to the cosmological singularities studied by Barrow~\cite{sudden}, but on on a timelike or near-null surface, rather than a spacelike surface.

We close this Section by stressing the all the theories in this paper have the same Newtonian limit, but different predictions beyond the Newtonian limit, depending on the clipper used. To illustrate this point, in Appendix~\ref{Sec:radial} we briefly repeat most of the exercise in this Section with a radial (instead of a transverse) clipper.

\section{Towards phenomenology}\label{phenomenology}
The models we have just presented have the drawback that they can evade any experimental constraints simply by setting the non-dynamical $\Delta$ to zero wherever the test was carried out (since there is no pretension at universal gravitation). This valid criticism is counteracted by the fact that, reciprocally, if $\Delta\neq 0$, these models would imply inimitable signatures.

\subsection{Proof of concept for testability}
It is not difficult to devise stylized situations with the unmistakable imprint of $\Delta$ matter, which can be in the realm of {\it gedanken} experiments or not. For these tests to be generic (and not restricted to a given relativistic theory and its specific clipping) they should probe the subverted Newtonian limit only.  

As an example, consider a Newtonian binary system, where one observes the wobble of a luminous object due to a dark companion. If the dark companion is made of $\Delta$ matter with (active gravitational) mass $m_2$ and the visible object has mass $m_1$, then 
for circular orbits with radius $a_1$ we would have the Keplerian relation for the orbital frequency $\omega$:
\begin{equation}
    \omega^2 a_1^3=m_2
\end{equation}
whatever the value of $m_1$.
In contrast to this subverted Newtonian limit, the usual Newtonian theory abiding by momentum conservation displays:
\begin{equation}
    \omega^2 a_1^3=\frac{m_2^3}{(m_1+m_2)^2}.
\end{equation}
The point is that we can conceive situations where $m_1$, $a_1$ and $\omega$ are measurable, so we would know the difference between the 2 theories, particularly if $m_2\sim m_1$. 

We can complicate this setup as much as wanted (eliptic orbits, several objects, etc). A relation with the methods of~\cite{willpsr}  and other tests of momentum conservation is possible. However, as we stress below, the usefulness of PPN parameters ($\zeta_{2}$ in the case of~\cite{willpsr}) is questionable, as we will explain further below. Note that it is enough to remark that PSR 1913 +16 is not made of $\Delta$ matter for the PPN parameters to revert to General Relativity values in this theory. 

\subsection{Lensing (and $\gamma_{PPN}$)}
Beyond the Newtonian limit, different clippings have different predictions (so a radial clipping, as outlined in Appendix~\ref{Sec:radial}, leads to different results). 
In addition, an analysis along the lines of~\cite{KostasPN} (in the unrelated context of the krhonon) may be more useful than standard PPN parameters. 
It is easy to see that the basic assumptions leading to the PPN parameterization (see for example~\cite{MTW}) break down for this theory (which is metric, nonetheless). 
The PPN parameters vary in space and depend on the configuration of $\Delta$ and of other forms of matter. 
For example, for the solutions in Section~\ref{SStesttube} we could read off:
\begin{equation}
    \gamma_{PPN}(r)=-\frac{m(r)}{r\Phi(r)}
\end{equation}
with different functions associated with \eqref{halometric} and \eqref{BHhalometric}, and all PPN parameters reverting their General Relativity value where $\Delta=0$. Any lensing experiment is far from local and would probe $\gamma_{PPN}$ along a light path, so the usefulness of the concept is dubious. 

It makes more sense to consider directly the lens equation following from integrating the geodesic equation for the various metrics obtained for these models. For example, for metrics \eqref{halometric}
and \eqref{BHhalometric}, with $A$ and $B$ defined from $ds^2=-A^2dt^2+B^2 dr^2 +r^2 d\Omega^2$, one would need to integrate:
\begin{equation}
    u'^2=\frac{E^2}{L^2A^2B^2}-\frac{u^2}{B^2}
\end{equation}
(with $u=1/r$, $'=d/d\phi$, and standard constants of motion $E$ and $L$), for light rays that penetrate the halo. For those that do not, we get the standard 
General Relativity formulae. 

Once again, this simplistic set up can be complicated as much as needed to allow comparison with observations. The effect on Einstein's rings is interesting and will be investigated elsewhere.

\subsection{Gravity waves and strong gravity tests}
In the weak field limit (since gravity then satisfies superposition and the refilled theory is equivalent to General Relativity with a very unusual non-dynamical medium) the gravitational wave solutions remain the same. Beyond the weak field limit there might be interesting effects for gravitational waves propagating through a $\Delta$ halo (possibly related to other work on gravity waves propagating through a medium, e.g.~\cite{Gravwavesmatter}). 

Note that within the refilled picture, the $\Delta$ matter must react to the wave to remain unmoved by its tidal stresses, so a wave of Stueckelberg stresses is expected and could backreact on the wave creating an interesting resonance. This could happen in particular near the horizon of a black hole with a $\Delta$ halo in attendance, as studied in Section~\ref{BHhalo}, where the effects are already non-linear. More generally, there could be an interesting impact on strong gravity phenomenon, for example regarding  black hole formation.

\section{Hamiltonian views}\label{HamSec}
There is a close connection between painted-on dark matter and the Hamiltonian framework. As pointed out in different forms in~\cite{shinji,geoCDM,nongeoCDM,StueckelDM,Barvinsky,kaplan1,kaplan2} the ``effective dark matter fluid'' can be seen as the legacy effect of a violation of the Hamiltonian constraint. This may be variously labeled as a violation of diffeomorphism invariance or a (local) Lorentz invariance violation (LIV; see also related literature in flat~\cite{DSRGianni,DSRLee} and curved~\cite{rainbow} space-time). 
This violation can be historical (for example due to past evolution~\cite{evol} or any other past interaction with a foliation~\cite{geoCDM,nongeoCDM}),
or else a high-energy effect~\cite{shinji,HL}. A combination of the two is also possible. The question then is what to do with the non-vanishing Hamiltonian once the LIV effects cease, if they do cease.

If the underlying theory satisfies the Dirac hypersurface deformation algebra (HDA; \cite{Dirac,DiracCanadian,Thiemann}) after symmetry restoration, then it is possible to fully restore the symmetry, with the effects of the symmetry breaking phase limited to a putative effective matter component, the total satisfying the Hamiltonian constraint {\it and} Dirac's HDA. This matter component is a dust ``CDM'' fluid if the preferred foliation is geodesic (e.g.~\cite{shinji,geoCDM}), and something more complicated for non-geodesic foliations (e.g.~\cite{nongeoCDM}).  

But what if the damage done to 4D diffeomorphism invariance 
is permanent as is the case envisaged towards the end of~\cite{MachianCDM}, where the residual Hamiltonian gets frozen-in in, as a time-independent $\Delta$? 

\subsection{LIV and let LIV in the Hamiltonian perspective}

If this is the case, then it is clear that the algebra of constraints must be modified in such regions. Recall~\cite{Dirac,DiracCanadian,Thiemann} that the Dirac HDA 
results from the extended Hamiltonian splitting as ${\cal H}_E=N{\cal H}+N^i{\cal H}_i$, with algebra:
\begin{align}
    \{H_i(f^i), H_j(g^j)\}&= H_i([\vec{f},\vec{g}]^i)\label{smearhihi}\\
    \{H_i(f^i), H(f)\}&= H(f^i\partial _if)\label{smearhih0}\\
    \{ H(f), H(f)\}&= H_i(h^{ij}(f\partial_j g- g\partial_j f)), \label{smearh0h0}
\end{align}
for smeared Hamiltonian $H(f)=\int d^3 y\,  f(y) {\cal H}(y)$ and momentum $H_i(f^i)=\int d^3 y\,  f^i (y){\cal H}_i(y)$ constraints. The HDA is equivalent to 4D diffeomorphism invariance.  Restoration of diffeomorphism invariance in the presence of a non-vanishing Hamiltonian (and possibly momentum~\cite{geoCDM,nongeoCDM}) then amounts to evolving the Hamiltonian and momentum according to this algebra, and then redefining the total Hamiltonian (and possibly momentum) absorbing the non-vanishing values into a fluid so that the total vanishes~\cite{geoCDM,nongeoCDM}. This is equivalent to  
the conservation 
of the Stueckelberg fields $S^{\mu\nu}$ defined in \eqref{Smunu}. Indeed, as proved in~\cite{nongeoCDM}, we have:
\begin{widetext}
\begin{align}
 n_\mu\nabla_\nu S^{\mu\nu}=0&\iff   \dot {\cal H}=\{{\cal H},{\bf H}\}_{HDA}=\partial_i(N^i {\cal H})+\partial_i ({\cal H}^i N)+ {\cal H}^i \partial_i N , \label{equiv1}\\
  h_{\alpha\mu}\nabla_\nu S^{\mu\nu}=0&\iff   
    \dot {\cal H}_i=\{{\cal H}_i,{\bf H}\}_{HDA}={\cal H}\, \partial_i N +\partial_j(N^j {\cal H}_i)  + {\cal H}_j\partial _i N^j, \label{equiv2}
\end{align}
\end{widetext}
with the total Hamiltonian defining the time evolution given by:
\begin{align}
     {\bf H}&=\int_{\Sigma_t} d^3 x\, {\cal H}_E(x).
\end{align}
From inspecting these equations, 
it is therefore clear that for a clipped theory 
with a non-conserved stress energy tensor associate with a dust-like $\Delta$ matter frozen-in on a non-geodesic frame, the HDA must be broken. The last term in \eqref{equiv1} precludes the time-independence of $\Delta$.

In fact, such $\Delta(x)$ can be seen as a central charge in the Dirac HDA, as we now show. 

\subsection{Painted-on matter as a central charge in the Dirac HDA}

Let us consider the effect of a non-dynamical frozen-in (time-independent) legacy Hamiltonian $\Delta$ which by construction is only a function of $x$ on a 
preferred foliation $\Sigma_t$: 
\begin{equation}
    \Delta=\Delta(x). 
\end{equation}
In a theory in which we recover the normal matter and gravity Lagrangian, we also recover the HDA for the sum of their Hamiltonians and momenta, represented by $GM$ in the expression for the extended Hamiltonian:
\begin{align}
    {\cal H}_E&=N{\cal H}+N^i{\cal H}_i\nn\\
    &={\cal H}_{GME}+N\Delta=N({\cal H}_{GM}+\Delta)+N^i{\cal H}^{GM}_{i}.
\end{align}
Hence the total Hamiltonian (including $\Delta$) and the momentum cannot satisfy the Dirac HDA. With generic smearing functions we must have instead:
\begin{align}
    \{H_i(f^i), H_j(g^j)\}&= H_i([\vec{f},\vec{g}]^i)\label{smearhihi}\\
    \{H_i(f^i), H(f)\}&= (H-\Delta)(f^i\partial _if)\label{smearhih0}\\
    \{ H(f), H(f)\}&= H_i(h^{ij}(f\partial_j g- g\partial_j f)), \label{smearh0h0}
\end{align}
implying on-shell relations (once one varies with respect to $N$ and $N^i$):
\begin{align}
    \{H_i(f^i), H_j(g^j)\}&\approx 0 \\
    \{H_i(f^i), H(f)\}&\approx -\int d^3x\, \Delta(x)f^i\partial _if\\
    \{ H(f), H(g)\}&\approx 0 . 
\end{align}
Alternatively the unsmeared algebra can be written as~\cite{Dirac,DiracCanadian,Thiemann}:
\begin{align}
    \{{\cal H}_i(x), {\cal H}_j(y)\}&={\cal H}_i(y)\delta_{,j}(x,y)+{\cal H}_j(x)\delta_{,i}(x,y)\label{hihi}\\
    \{{\cal H}_i(x), {\cal H}(y)\}&=({\cal H}(x)-\Delta(x))\delta_{,i}(x,y)\label{hih0}\\
    \{{\cal H}(x), {\cal H}(y)\}&=h^{ij}({\cal H}_i(x)\delta_{,j}(x,y)+{\cal H}_i(y)\delta_{,j}(x,y))\label{h0h0}
\end{align}
where $\delta_{,i}(x,y)=\frac{\partial}{\partial x^i}\delta(x,y)$, and delta is a bidensity of weight zero (i.e. a scalar) in the first argument and of weight one in the second~\cite{Isham}.   
Smeared or not, the relevant Poisson bracket is the one involving the Hamiltonian and the momentum, with on-shell anomalous result:
\begin{align}\label{secondclass}
     \{H_i(f^i), H(N)\}&\approx -\int d^3x\, \Delta(x)f^i\partial _iN\nn\\
      \{{\cal H}_i(x), {\cal H}(y)\}&\approx-\Delta (x) \frac{\partial}{\partial x^i}\delta(x,y).
\end{align}

In the face of this result, 
going through the Dirac procedure described in~\cite{Dirac} offers two major clarifications. These result from forks in the procedure, which arise not because the procedure is ambiguous, but because the way we laid out the theory has several options which have to be clarified, leading to such forks. Some of these possibilities do not serve the purpose of this paper.

First, as already stressed at the start of this paper, $\Delta$ must be non-dynamical, which in the canonical framework means it must not be seen as a phase space variable (with a conjugate momentum $\Pi_\Delta$ which does not appear anywhere). Otherwise, Dirac's procedure implies $\Delta=0$ as a secondary constraint, and we do not want that. Therefore, in the canonical formalism $\Delta(x)$ should be seen as a non-dynamical central charge in the Dirac HDA. We can keep it general or not, since it is an external, given function. With this proviso,  Dirac's HDA therefore contains second class constraints.

\subsection{Clipping as a Dirac bracket solution to the second class constraints}

The second clarification provided by Dirac's recipes~\cite{Dirac} is that we should not eliminate the second class constraints solely by restricting the Lagrange multipliers $N$ and $N^i$. Second class constraints may be solved by imposing conditions upon the Lagrange multipliers; in our case these could be obtained from the secondary constraints: 
\begin{align}
\dot {\cal H}=    \{ {\cal H},\mathbf{H}_E \}&\approx -\partial_i(N^i\Delta)=0\\
  \dot {\cal H}_i=  \{ {\cal H}_i,\mathbf{H}_E \}&\approx -\Delta\partial_iN =0
\end{align}
implying that wherever $\Delta\neq 0$ one must have:
\begin{align}
    \partial_i N&=0\label{lagred1}\\
    \partial_i(N^i\Delta)&=0.\label{lagred2}
\end{align}
But this is not what we want, since it would set to zero the source terms in \eqref{noncon1} and \eqref{noncon2}, which are the point of this paper. Indeed, 
imposing \eqref{lagred1} and \eqref{lagred2} implies a geodesic $\Sigma_t$ and a conserved dust fluid. This is old news: it is the IR limit of projectable HL theory~\cite{shinji} (with violations in the UV as described in~\cite{HLPaolo}), or the dust fluid appearing in the the evolution scenarios in~\cite{geoCDM} and also in~\cite{Barvinsky,kaplan1,kaplan2}.

We want the possibility of a non-conserved dust fluid in a non-geodesic $\Sigma_t$. The Dirac bracket~\cite{Dirac} offers a way in, and is equivalent to a clipping procedure as described in Sections~\ref{Sec:clippersGen} and \ref{Sec:clippers}. This can be seen appealing to the Maskawa–Nakajima theorem \cite{maskawa} which states that it is always possible to find a canonical transformation such that the second-class constraints take the form $q_r = p_r = 0$ for some of the new canonical pairs of variables (with the non-closing algebra reduced to $\{q_r,p_s\}=\delta_{rs}$). Then, the Dirac bracket is equivalent to the Poisson bracket evaluated with respect to all variables except these, which are to be ignored. This is equivalent to clipping out these variables. 

The number of pairs of second class constraints equals the number of canonical pairs to be clipped, and so the number of spatial ($h_{ij}$) metric components to be clipped. In contrast, if we eliminate the Lagrange multipliers, the number of second class constraints equals the number of multipliers eliminated. A hybrid approach is also possible (as advocated in Section~\ref{Sec:clippers}; see comments around \eqref{diffreduct}). The number of second class constraints could also depend on any conditions imposed on $\Delta$. 

Face value, the implication is that it should be possible to clip a single spatial metric variable in our problem, and this would be the solution provided by the Dirac bracket, but we leave this exercise to future work. Notice also that the procedure in Section~\ref{Sec:clippersGen} is more general, and allows for clippings that do not emerge from the Dirac bracket, as well as for over-clipping.





\section{Concluding appraisal}
We close with a general appraisal. 
The approach in this paper should be construed as a reaction against the one-explanation-fits-all perspective of many dark matter proponents. The object lesson of Uranus and Mercury anomalies in XIX century solar system theory is relevant here. Nowadays it does not offend anyone's Occam razor sensitivities that these anomalies were resolved by such disparate explanations as a conceptualized ``Neptune'' (i.e., as it was before it was actually sighted, and so an early example of ``dark matter'') and General Relativity (a glaring example of ``modified gravity'' in this context). Occam's razor would call for a Vulcanus in Mercury's case and, facetiously, for a new theory of gravity for the anomalies of Uranus. Likewise, the incidental shortcomings of the dark matter paradigm (swept under various carpets, such as bias, fuzzy dark matter and other epicycles) perhaps should be interpreted as evidence that dark matter is the correct explanation in some cases, but not in others; or that at least in some situations what we call dark matter might be a proxy for something entirely different in nature. 

In this paper we embraced the unorthodox notion of ``painted-on'' dark matter: something which is not matter at all but behaves like an esoteric dust  encrusted in a preferred foliation, which may be non-geodesic so that such ``dust'' violates conservation laws, and is a signature of an aether wind. Its theoretical origins may lie in violations of the Hamiltonian constraint, but in contrast with~\cite{shinji,geoCDM,nongeoCDM,StueckelDM,Barvinsky,kaplan1,kaplan2}, by freezing the leftover Hamiltonian, we are  entertaining the possibility that diffeomorphism invariance is never recovered in such regions, leading to violations of local energy-momentum conservation. A reverse analogy with topological defects can be drawn here. Topological defects arise when symmetry breaking takes place at low temperatures, the defect representing a region where the field remained stuck in the high temperature symmetric phase. We have the reverse setup: at low energies (or at late times) we restore the very fundamental diffeomorphism symmetry, which is broken at high energies (or in the early Universe). We then  envisage the possibility that defects linger on, where (for non-topological reasons) the broken phase survives, and that this might be part of the ``dark matter''.

Regardless of this motivation, it is important to stress that ``painted-on'' dark matter was in fact a colloquial term used by a few founding fathers of the dark matter paradigm, to embody wishful thinking for desirable behaviour which traditional dark matter failed to exhibit. Clearly the idea is terrible for explaining what conventional dark matter does in the linear regime. Ditto regarding the many straightforward successes of dark matter in the non-linear regime, such as lensing in its many guises. But just as with Neptune and Vulcanus, it might be fruitful to reject reductionism and accept that an entirely different concept might apply where the standard dark matter does not fare so well.

In setting up this theory and its solutions, 
we took inspiration from galactic halos, but we did not want to be slaves of phenomenology: the short-sighted drive to fit the data by forceful models at all costs. We do not explain the Tully-Fisher relation, the role of an acceleration scale in triggering anomalies, and many other details of galactic scale phenomenology. Given galaxies' peculiar velocities, $\Sigma_t$ would also have to be dissociated from the cosmological frame, possibly creating problems. We do not apologize for these shortcomings. This paper is not only anti-reductionist but anti-phenomenologist. A good idea with incomplete phenomenology is better than a shallow recipe for fitting all the data. 

But is our proposal a good idea? In closing, we accept that further conceptual work is warranted on the problems associated with coupling gravity to a non-conserved source in a relativistic theory. In this paper we suggested a solution: {\it clipping}, which can take the form of a Dirac bracket in the Hamiltonian formulation. But the Bianchi identities are so powerful that they always allow for {\it refilling} of the clipped equations with suitable (Stueckelberg) stresses. Non-conservation can then masquerade as conservation supplemented by appropriate stresses, but, as we showed, by making the equations for these stresses suitably ludicrous, we arrive at a possible definition of broken diffeomorphism invariance. Is this radical enough? Is there a more fundamental construction for a theory with non-conserved sources?



\section{Acknowledgments}
We thank Dick Bond, Bianca Dittrich and Kostas Skordis for help with this paper. JM is grateful to the participants of the Caribbean Futures Of Science Symposium (where this work was first presented) for equal measures of encouragement and acerbic criticism.  RI was supported by a Bell-Burnell Fellowship and JM partly supported by STFC Consolidated Grant ST/T000791/1.

\appendix 

\section{A brief outline of radial clipping}\label{Sec:radial}

As an example of how different the theory could be with an alternative clipping prescription, in this Appendix we consider a radial clipping, whilst fixing the transverse metric, so that the spherically symmetric equations are reduced to:
\begin{align}
    e^{-2\lambda} \left[ \frac{2}{r} \lambda' - \frac{1}{r^2} \right] + \frac{1}{r^2} &= 8 \pi (\rho_\Delta+\rho_M),\label{G00}\\
    e^{-2\lambda} \left[ \Phi'' + \Phi'^2 - \Phi' \lambda' + \frac{\Phi'}{r} - \frac{\lambda'}{r} \right] &= 8 \pi p_M,\label{Gangular}\\
    -\Phi'(\rho_M+p_M) &=p_M'\label{MCons}
\end{align}
as opposed to the system \eqref{G00T}, \eqref{GrrT} and \eqref{MConsT}. The integration \eqref{lambdamrT} and \eqref{mprimeT} remains the same, but the equation for $\Phi$ is different. For a critical halo we have the same expression for $\lambda$ but $\Phi$
loses sight of $\Delta$:
\begin{equation}
    \Phi'' + \Phi'^2 + \frac{\Phi'}{r} = 0
\end{equation}
with solution:
\begin{equation}
    \Phi=\Phi_0 + \ln[\ln r/r_0],
\end{equation}
where $\Phi_0$ is to be determined by the boundary conditions. 
Instead of \eqref{halometric}, the critical halo metric is:
\begin{equation}\label{halometricR}
    ds^2=-e^{2\Phi_0}\ln^2\frac{r}{r_0} dt^2+\frac{dr^2}{1-2m_0}+r^2d\Omega^2.
\end{equation}
The Newtonian limit of this solution is identical to that obtained with a transverse clipping (in particular it creates flat rotation curves), but the PN and strong gravity predictions outlined in Section~\ref{phenomenology}  are different. 

The refilling Stueckelberg stresses described in Section~\ref{refillSS} are also different. We can define an anisotropic radial stress completion giving us back \eqref{GrrT} with a right hand side modified by
\begin{equation}\label{preff}
p_M\rightarrow p_M+p_{\Delta r}    
\end{equation}
such that:
 \begin{equation}\label{preposterous}
     \frac{1}{r^2}(r^2 p_{\Delta r})'= -\Phi'(\rho_\Delta+p_{\Delta r}).
 \end{equation}
This refilling is to be contrated with  \eqref{peffT} and \eqref{PeffeqT} for a transverse clipping. Then, just as before, we could find the $\Phi$ field integrating the usual radial equation with this radial stress added on. 
Again this is hiding the fact that $p_{\Delta r}$ is not related to $\Delta$ via any equation of state or local dynamics minimally coupled to gravity.


\begin{thebibliography}{99}

\bibitem{fuzzy}
W.~Hu, R.~Barkana and A.~Gruzinov,
Phys. Rev. Lett. \textbf{85}, 1158-1161 (2000)
doi:10.1103/PhysRevLett.85.1158

\bibitem{triaxial}P. M. W. Kalberla 2003 ApJ 588 805.

\bibitem{HL}
P.~Horava,
Phys. Rev. D \textbf{79}, 084008 (2009)
doi:10.1103/PhysRevD.79.084008
[arXiv:0901.3775 [hep-th]].

\bibitem{shinji}
S.~Mukohyama,
Phys. Rev. D \textbf{80}, 064005 (2009)
doi:10.1103/PhysRevD.80.064005
[arXiv:0905.3563 [hep-th]].





\bibitem{geoCDM} 
J.~Magueijo,
Phys. Rev. D \textbf{109}, no.12, 124026 (2024)
doi:10.1103/PhysRevD.109.124026
[arXiv:2404.15809 [hep-th]].

\bibitem{nongeoCDM}
J.~Magueijo,
Phys. Rev. D \textbf{110}, no.8, 084050 (2024)
doi:10.1103/PhysRevD.110.084050
[arXiv:2406.17428 [gr-qc]].

\bibitem{StueckelDM}
R.~Casadio, L.~Chataignier, A.~Y.~Kamenshchik, F.~G.~Pedro, A.~Tronconi and G.~Venturi,
[arXiv:2402.12437 [gr-qc]].

\bibitem{Barvinsky}
A.~O.~Barvinsky and A.~Y.~Kamenshchik,
Phys. Lett. B \textbf{774}, 59-63 (2017)
doi:10.1016/j.physletb.2017.09.045
[arXiv:1705.09470 [gr-qc]].


\bibitem{kaplan1}
L.~Del Grosso, D.~E.~Kaplan, T.~Melia, V.~Poulin, S.~Rajendran and T.~L.~Smith,
[arXiv:2405.06374 [hep-ph]].

\bibitem{kaplan2}
D.~E.~Kaplan, T.~Melia and S.~Rajendran,
[arXiv:2305.01798 [hep-th]].

\bibitem{mimetic}
A.~H.~Chamseddine and V.~Mukhanov,
JHEP \textbf{11}, 135 (2013)
doi:10.1007/JHEP11(2013)135
[arXiv:1308.5410 [astro-ph.CO]].

\bibitem{MachianCDM}
J.~Magueijo,
Phys. Lett. B \textbf{858}, 139001 (2024)
doi:10.1016/j.physletb.2024.139001
[arXiv:2312.07597 [hep-th]].





\bibitem{unimod} M.~Henneaux and C.~Teitelboim, ``The cosmological constant and general covariance,'' {\em Phys.\ Lett.\ B} {\bf 222} (1989), 195--199.

\bibitem{unimod1}
W. G. Unruh, Phys. Rev. {\bf D40}, 1048 (1989);  K.~V.~Kucha\v{r}, ``Does an unspecified cosmological constant solve the problem of time in quantum gravity?,'' {\em Phys.\ Rev.\ D} {\bf 43} (1991), 3332--3344.


\bibitem{UnimodLee1}
L.~Smolin, ``Quantization of unimodular gravity and the cosmological constant problems,'' {\em Phys.\ Rev.\ D} {\bf 80} (2009), 084003, arXive: 0904.4841. 

\bibitem{alan} A. Daughton, J. Louko, and R. D. Sorkin, ``Instantons and unitarity in quantum cosmology with fixed four-volume,'' {\em Phys.\ Rev.\ D} {\bf 58}, 084008 (1998).

\bibitem{daughton} A. Daughton, J. Louko, and R. D. Sorkin, ``Initial conditions and unitarity in unimodular quantum cosmology,'' [gr-qc/9305016].

\bibitem{sorkin1} R. D. Sorkin, ``Role of time in the sum-over-histories framework for gravity,'' {\em Int J Theor Phys} {\bf 33}, 523–534 (1994). https://doi.org/10.1007/BF00670514

\bibitem{sorkin2} R. D. Sorkin, ``Forks in the road, on the way to quantum gravity,'' {\em Int J Theor Phys} {\bf 36}, 2759–2781 (1997). https://doi.org/10.1007/BF02435709


\bibitem{Bombelli}
L.~Bombelli, W.~E.~Couch and R.~J.~Torrence,
Phys. Rev. D \textbf{44}, 2589-2592 (1991)
doi:10.1103/PhysRevD.44.2589

\bibitem{UnimodLee2}
L.~Smolin,
Phys. Rev. D \textbf{84}, 044047 (2011)
doi:10.1103/PhysRevD.84.044047
[arXiv:1008.1759 [hep-th]].

\bibitem{Saltas}
A.~Padilla and I.~D.~Saltas,
Eur. Phys. J. C \textbf{75}, no.11, 561 (2015)
doi:10.1140/epjc/s10052-015-3767-0
[arXiv:1409.3573 [gr-qc]].

\bibitem{Fried}
K.O.Friedrichs, Math. Ann 98, 566 (1928).

\bibitem{BHevol}
J.~Magueijo,
Phys. Rev. D \textbf{109}, no.4, 044034 (2024)
doi:10.1103/PhysRevD.109.044034
[arXiv:2310.11929 [hep-th]].

\bibitem{sudden}
J.~D.~Barrow,
Class. Quant. Grav. \textbf{21}, L79-L82 (2004)
doi:10.1088/0264-9381/21/11/L03
[arXiv:gr-qc/0403084 [gr-qc]].

\bibitem{evol}
J.~Magueijo,
Phys. Rev. D \textbf{108}, no.10, 103514 (2023)
doi:10.1103/PhysRevD.108.103514
[arXiv:2306.08390 [hep-th]].

\bibitem{willpsr}
C. Will, Ap. J. Letts. {\textbf 393}, L59 (1992).



\bibitem{MTW} C.~Misner, K.~Thorne and J.A.Wheeler, Gravitation,  W. H. Freeman, 1973.

\bibitem{KostasPN}
L.~Blanchet and C.~Skordis,
JCAP \textbf{11}, 040 (2024)
doi:10.1088/1475-7516/2024/11/040
[arXiv:2404.06584 [gr-qc]].

\bibitem{ADMReview}
R.~Jha,
``Introduction to Hamiltonian formulation of general relativity and homogeneous cosmologies,''
SciPost Phys. Lect. Notes \textbf{73}, 1 (2023); 
[arXiv:2204.03537 [gr-qc]].


\bibitem{Gravwavesmatter}
N.~T.~Bishop, P.~J.~van der Walt and M.~Naidoo,
Phys. Rev. D \textbf{106}, no.8, 084018 (2022)
doi:10.1103/PhysRevD.106.084018
[arXiv:2206.15103 [gr-qc]].


\bibitem{DSRGianni}
G.~Amelino-Camelia,
Int. J. Mod. Phys. D \textbf{11}, 35-60 (2002)
doi:10.1142/S0218271802001330
[arXiv:gr-qc/0012051 [gr-qc]].

\bibitem{DSRLee}
J.~Magueijo and L.~Smolin,
Phys. Rev. Lett. \textbf{88}, 190403 (2002)
doi:10.1103/PhysRevLett.88.190403
[arXiv:hep-th/0112090 [hep-th]].

\bibitem{rainbow}
J.~Magueijo and L.~Smolin,
Class. Quant. Grav. \textbf{21}, 1725-1736 (2004)
doi:10.1088/0264-9381/21/7/001
[arXiv:gr-qc/0305055 [gr-qc]].





\bibitem{Dirac}P.~Dirac, “Lectures on Quantum Mechanics”, Belfer Graduate School of Science, Yeshiva
University Press, New York, 1964. 

\bibitem{DiracCanadian}P.~A.~M.~Dirac,
Can. J. Math. \textbf{2}, 129-148 (1950).

\bibitem{Thiemann}T. Thiemann, ``Modern canonical quantum general relativity'', Cambridge Monographs on Mathematical Physics, CUP, Cambridge, 2008. 



\bibitem{Isham}
C.~J.~Isham,
NATO Sci. Ser. C \textbf{409}, 157-287 (1993)
[arXiv:gr-qc/9210011 [gr-qc]].

\bibitem{HLPaolo}
P.~M.~Bassani, J.~Magueijo and S.~Mukohyama,
[arXiv:2408.03793 [gr-qc]].


\bibitem{maskawa} T. Maskawa and H. Nakajima,
Progress of Theoretical Physics, 56:1295–1309, 1976.



\end{thebibliography}
\end{document}